# Quantum-Based Resilient Routing in Networks: Minimizing Latency Under Dual-Link Failures

**Maher Harb[1], Nader Foroughi[1], Matt Stehman[1], Bob Lutz[1], Nati Erez[2], Erik Garcell[2]**

[1] Comcast Corporation, Philadelphia, PA 19103 USA
[2] Classiq Technologies, 3 Daniel Frisch Street, Tel Aviv-Yafo, 6473104, Israel

Corresponding author: Maher Harb (email: maher.harb@gmail.com).

**Abstract.** Network optimization problems represent large combinatorial search spaces that grow exponentially with network size, making them computationally intensive to solve. This paper addresses the latency-resilient Layer 3 routing optimization problem in telecommunications networks with predefined Layer 1 optical links. We formulate this problem as a graph-based optimization problem with the objective of minimizing latency, creating vertex-disjoint paths from each site to the internet backbone, and maximizing overall resiliency by limiting the impact of dual-link failures. By framing the problem as finding two disjoint shortest paths, coupled together with a resiliency component to the objective function, we establish a single formulation to produce optimal path design. The mathematical formulation was adapted to solve the problem using quantum approximate optimization algorithm (QAOA) executed over both quantum simulator and quantum hardware. QAOA was tested on a toy graph topology with 5 vertices and 7 edges and considering two limiting scenarios respectively representing independent (uncorrelated) link failures and highly correlated failure for one pair of edges. Both explored scenarios produced the optimal network design—corresponding to the valid solution with highest frequency of occurrence and minimum energy state, hence, validating the proposed formulation for optimizing Layer 3 routing on quantum systems of the future.

## 1. Introduction

Quantum algorithms present a promising opportunity for tackling complex network optimization challenges that can be modeled as graph problems. Initially proposed by Farhi *et al.* in 2014, the Quantum Approximate Optimization Algorithm (QAOA) is now a foundational approach in quantum computing for solving combinatorial optimization problems, such as maximum cut, graph coloring, and network flow [1-4]. To address the shortest path problem, researchers have explored quantum walk-based algorithms that leverage quantum superposition to explore multiple paths in parallel, potentially offering speedups over classical Dijkstra's algorithm for specific graph topologies [5-8]. Khadiev and Safina, for instance, introduced a quantum algorithm designed for finding the shortest path in directed acyclic graphs with time complexity $O(\sqrt{|V||E|} \cdot \log |V|)$, where |V| and |E| respectively represent a graph's number of vertices and number of edges [8]. Additionally, quantum annealing approaches have been investigated for solving traffic flow and shortest-path-type problems using D-Wave specialized quantum annealers, by encoding optimization objectives and constraints into quadratic unconstrained binary optimization (QUBO) formulations [9-13].

Furthermore, there has been equal interest in hybrid quantum-classical approaches that leverage variational quantum algorithms to solve network optimization scenarios. For instance, for network flow problems, including maximum flow and minimum cost flow, researchers have developed quantum formulations using amplitude amplification and quantum linear system solvers (example, the HHL algorithm—Harrow, Hassidim, and Lloyd), though practical implementations remain limited by current hardware constraints such as qubit connectivity and gate fidelity [14-16]. Quantum machine learning approaches have also been applied to learn optimal routing policies in dynamic networks [17]. However, significant challenges persist, including the overhead required to encode classical graph data into quantum states, the depth of quantum circuits needed for problems of practical network size, and the question of whether quantum advantage can be achieved for these problems given that many classical approximation algorithms already provide efficient solutions. Current research suggests that the implementations of quantum annealing in optimization are limited in size and not yet scalable to real-world situations [12]. In contrast, recent studies on traffic flow optimization show that hybrid quantum annealing achieves solutions that are near-optimal and within 1% of classical solvers [13]. Thus, empirical demonstrations of practical quantum advantage in network optimization remain an active area of investigation.

The practical challenges identified above—particularly in achieving quantum advantage for realistic problem sizes—become even more evident when considering modern telecommunications networks. These networks increasingly employ meshy topologies where multiple candidate paths exist between service endpoints. A critical operational requirement in routing design over meshy topologies is to provision two vertex-disjoint paths between each secondary site and backbone hubs: one primary and one backup path. These paths must simultaneously minimize end-to-end latency for performance and minimize exposure to correlated link failures for resilience. When failure probabilities are not precisely known, operators typically assume uniform failure distributions across links, making the minimization of shared failure exposure between path pairs a key design objective. This multi-objective combinatorial problem—balancing speed and resilience while maintaining path redundancy—is computationally challenging, particularly as network size scales.

Yet despite the theoretical promise of quantum approaches for network optimization, their application to specific telecommunications challenges remains largely unexplored. This paper addresses the latency-resilient Layer 3 (L3) path mapping problem for meshy network topologies. We present two solver realizations of the same model. The first realization is an integer linear program (ILP) formulation that enforces flow conservation and path redundancy while optimizing a composite objective function that includes the latency and resiliency components. The second realization is an equivalent QUBO formulation that maps directly to a cost Hamiltonian for solving the algorithm on quantum hardware. The key contributions of this work are:

- A unified multi-objective formulation for computing a pair of disjoint paths that jointly minimizes end-to-end latency and failure-impact under independent (uncorrelated) and dependent (correlated) dual-link failure models.
- A novel resiliency metric based on pairwise joint-failure probabilities that penalizes cross-path link exposure, enabling solutions that systematically avoid high-risk link combinations.
- Dual solver implementations: (a) an ILP formulation suitable for classical discrete optimization solvers, and (b) an equivalent QUBO formulation enabling solution via QAOA on near-term quantum hardware.
- Comparative evaluation demonstrating how the model selects low-latency routes while reducing joint-failure exposure in representative meshy topologies.

The paper is organized as follows. Section II introduces the problem on a toy graph topology. Section III develops the mathematical formulation using a set of linear equations and constraints. Section IV comments on the computational complexity of the problem. Section V adapts the mathematical formulation to the QUBO form. Section VI presents the QAOA approach. Section VII presents and discusses the results of QAOA executed on both a quantum simulator and quantum hardware. Section VIII concludes with remarks on the prospect of executing QAOA for real network topologies with typical scale of ~100 vertices and edges.

## 2. Problem Statement

The latency-resilient L3 routing problem is best described in relation to the graph, $G = (V, E)$, shown in Fig. 1, which represents hypothetical Layer 1 (L1) topology. Each blue-colored (light) vertex is a secondary site. The two red-colored (dark) vertices—labeled as T1 & T2—are terminal sites that connect to the internet backbone. The labels on the edges correspond to the latency of the optical links in arbitrary units. In addition, each link is assumed to have a probability of failure (not explicitly shown on graph).

Formally, we seek two vertex-disjoint paths for each secondary site respectively to terminal sites T1 & T2 as destinations, with the objective of minimizing total edge cost (latency). In addition, given knowledge of the joint probability of failure for each pair of optical links in the network, the ideal design should correspond to the one which minimizes the impact of two failed links—in terms of number of isolated subscribers. Example, for source site S2, the objective is to find two disjoint shortest paths for source-destination pairs (S2, T1) and (S2, T2). Note that vertex-disjoint is a stricter requirement from the perspective of resiliency than edge-disjoint: it does not allow the two paths to have any common vertices aside from the source vertex S2. By extension, vertex-disjoint paths are also edge-disjoint. Furthermore, a solution in which the two disjoint paths include two highly correlated links should be less desired, even if it corresponds to the minimum latency solution.

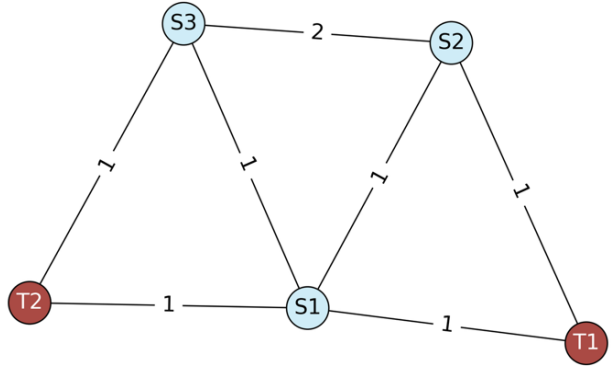

**Figure 1. Illustrative Layer-1 (L1) Network Topology.** Blue (light) circles are secondary sites (each feeds a designated market), and red (dark) circles (T1 & T2) are terminal sites that also serve as hubs to the internet backbone. The labels on edges represent the latency of the optical link in arb. units. Hence, this L1 topology is represented as a graph with 5 vertices and 7 edges. Each secondary site must have two vertex-disjoint paths to T1 and T2.

## 3. Mathematical Formulation of Problem

In this section we express the L3 routing problem using an ILP formulation in which objective function and constraints are coded as linear expressions. Throughout the problem modeling, we assume the graph to be undirected and to contain only non-negative edge weights. First, we start with the following definition of notation (expanding on the notation in [11]):

- V: set of graph vertices (includes both the secondary sites and the two terminal sites/backbone hubs)
- E: set of graph edges (L1 optical fiber links)
- A vertex is referred to by its index in set V
- An edge is referred to by the two vertices it connects; example, edge $(i, j)$ connects vertices $i$ and $j$
- $s$: a specific source/start vertex; always a secondary site. Note that $s$ is fixed—the problem is solved independently for each source site
- $t_1, t_2$: the two terminal vertices (backbone hubs)
- $T = \{t_1, t_2\}$: Set of terminal vertices
- $P^{(s,l)}$: A path that connects source vertex s with terminal vertex $l$
- $x_i^{(s,l)}$: binary decision variable indicating whether vertex $i$ is included in path $P^{(s,l)}$
- $d_s$: number of subscribers connected to source $s$
- $x_{i,j}^{(s,l)}$: binary decision variable indicating whether edge $(i,j)$ is included in path $P^{(s,l)}$
- $c_{ij}$: latency cost of using edge $(i, j)$
- $p(ij, mn)$: joint probability of failure for edges $(i, j)$ and $(m, n)$

Creating optimal paths for a given source vertex $s$ is expressed through the following objective:

$$\min f(x) + Bg(x), \tag{1}$$

where $f(x)$ and $g(x)$ respectively represent the latency and resiliency components of the objective function—to be defined, and $B$ is a scaling factor that can be regarded as a mechanism to normalize the units across the two components, and more importantly, adjust the trade-off between them. $f(x)$ is the total latency cost of edges used over the two paths, defined as:

$$f(x) = \sum_{l \in T} \sum_{i,j \in E} c_{i,j} x_{i,j}^{(s,l)}, \tag{2}$$

subject to $x_{i,j}^{(s,l)} \in \{0,1\}, \forall (i,j) \in E, \forall l \in T$. The following path connectivity constraint ensures that the chosen edges form a valid (connected with no loops) path from the source vertex to the destination vertex:

$$\sum_j x_{i,j}^{(s,l)} = \begin{cases} 1, & \text{if } i \in \{s, l\} \\ 2, & \text{if } i \in P^{(s,l)} \text{ and } i \notin \{s, l\}, \\ 0, & \text{if } i \notin P^{(s,l)} \end{cases} \tag{3}$$

where $i \in V$. The next constraint ensures that the two paths $P^{(s,t_1)}$ and $P^{(s,t_2)}$, starting from the same source $s$, are vertex-disjoint (i.e., they contain no common vertices other than the source vertex):

$$\sum_{l \in T} x_i^{(s,l)} \leq 1 \text{ if } i \in V \setminus \{s\}. \tag{4}$$

The resiliency component, $g(x)$, represents the impact of dual fiber link failures, and is defined as:

$$g(x) = d_s \sum_{i,j \in E} \sum_{m,n \in E} p(ij, mn) x_{i,j}^{(s,t1)} x_{m,n}^{(s,t2)}, \tag{5}$$

where demand $d_s$, the number of subscribers attached to the source vertex $s$, is assumed to be 1. Notice that choice of $d_s$ does not hold any impact on the solution beyond adjusting the value of scaling factor $B$. The way $g(x)$ is formulated and given a pair of edges $(i,j)$ & $(m,n)$, the argument of the summation will be non-zero only if each of the pair of edges is included in one of the two paths $P^{(s,t_1)}$ and $P^{(s,t_2)}$. This is because if both edges are part of the same path, then the other factor in the product will be zero, owing to the vertex-disjoint constraint expressed in (4). Therefore, the argument in (5) assumes either a value of $p(ij, mn)$ or zero.

As formulated, each one path from source $s$ requires number of binary decision variables equal to the number of edges and vertices. Thus, the required number of binary decision variables to model the problem is $2(|V| + |E|)$, resulting in 24 variables for the toy topology presented in Fig. 1. Moreover, even though the introduced formulation includes a quadratic term in (5), it can be easily adapted to a linear formulation by substituting $x_{i,j}^{(s,t1)} x_{m,n}^{(s,t2)}$ with a slack variable along with the necessary set of constrains. Such substitution comes at a steep expense of additional $|E|^2$ variables (adding another 49 variables for the considered graph topology).

Equations (1) to (5) guarantee the creation of valid disjoint paths (if permitted by graph topology), and while the objective function aims to minimize the combination of total path latency and failure impact. Parameter $B$ allows adjusting the trade-off between latency and resiliency. With $B = 0$, the problem simplifies into finding two disjoint shortest paths. On the other hand, by setting $B$ to a value that is much larger than the total sum of edge weights, the problem becomes an ordered optimization problem in which we minimize $f(x)$ subject to $g(x)$ being minimized.

## 4. Related problems and Complexity

So-called "disjoint path problems" and their computational complexity are well studied in mathematics, computer science and engineering literature (e.g., see [18] for a survey of results). Many of these problems can be phrased as follows: Given an undirected graph $G$ with source vertex $s$ and terminal vertex $t$, find internally vertex-disjoint paths $P_1$ and $P_2$ from $s$ to $t$ such that some non-negative objective function $\lambda(P_1, P_2)$ is minimized.

Whereas the arguments of $f(x)$ and $g(x)$ in Section III are binary vectors that encode paths indirectly, here $\lambda$ is a function of the paths themselves. We consider the domain of $\lambda$ to be the set of all pairs $(P_1, P_2)$ of paths from $s$ to $t$ such that $P_1$ and $P_2$ are internally vertex-disjoint, so that the problem is simply to find a global minimum of $\lambda$.

The complexity of this problem depends subtly on the function $\lambda$. For example, letting $|P|$ denote the number of edges in path $P$, the *min-sum* problem takes $\lambda(P_1, P_2) = |P_1| + |P_2|$ to be the total number of edges in $P_1$ and $P_2$. This problem can be solved in polynomial time using a suitable modification of Suurballe's algorithm [19] or a standard minimum-cost flow approach (e.g. successive shortest path). But the problem typically becomes more complex when $\lambda$ involves nonlinear functions of $|P_1|$ and $|P_2|$. For example, if $\lambda(P_1, P_2) = (|P_1|^p + |P_2|^p)^{1/p}$, then it can be shown that the problem is NP-hard for all $p > 1$. In the limit as $p \to \infty$, we obtain the *min-max* problem $\lambda(P_1, P_2) = \max(|P_1|, |P_2|)$, which is NP-hard as well [20]. The corresponding *min-min* problem, i.e. with $\lambda(P_1, P_2) = \min(|P_1|, |P_2|)$, is also known to be NP-hard [21].

The L3 latency-resilient routing problem can be represented to fit neatly in this framework while introducing a multiplicative objective that makes it novel among disjoint path problems. For this we introduce the objective function

$$\lambda(P_1, P_2) = (|P_1| + C) \cdot (|P_2| + C), \tag{6}$$

where $C \geq 0$ is an arbitrary constant. Following *min-sum*, *min-max*, and *min-min*, we call this the *min-prod* disjoint paths problem. Expanding the product in (6) reveals the relationship to the L3 problem; the term $C(|P_1| + |P_2|)$ represents the latency component $f(x)$ while the quadratic term $|P_1| \cdot |P_2|$ stands in for the dual link-failure resiliency component $g(x)$.

While it seems highly likely that *min-prod* is NP-hard, the problem permits no obvious reductions from known NP-hard problems in the area, including *min-min*, *min-max*, and the related problem of finding $P_1$ and $P_2$ such that $\min(|P_1|, |P_2|) \leq \Delta_1$ and $\max(|P_1|, |P_2|) \leq \Delta_2$ for given constants $0 < \Delta_1 \leq \Delta_2$ [22].

In general, disjoint path problems with objectives containing products of lengths do not seem to appear in the literature, and it is precisely this operation that makes reductions from the standard NP-hard problems difficult. We conjecture that *min-prod* is NP-hard but leave its complexity as an open question. Lastly, we refer to the Appendix for a formal proof that NP-hardness of *min-prod* implies NP-hardness of the latency-resilient L3 routing problem.

## 5. Formulation as Quadratic Unconstrained Binary Optimization

In this section we rewrite the ILP formulation as QUBO. The re-formulation into the QUBO form allows for the optimization to be run on quantum hardware where the binary decision variables (logical or algorithmic qubits) are mapped onto physical qubits. Ideally the mapping from logical to physical qubits is 1-1. However, hardware implementations may use more physical qubits than logical qubits, either due to topology restrictions (mainly in quantum annealers) or as required by the error correction mechanism. In any case, we will not be concerned with the actual mapping of the logical-to-physical qubits beyond the restriction it imposes on the scale of the problem.

The transformation from an objective function and set of constraints to QUBO is a standard process with some guiding principles on the approach outlined below:

- Binary Encoding: Each binary decision variable maps directly into a qubit. However, because QUBO is a single expression that inherently does not allow inequalities, additional slack variables may be needed to encode inequalities (see [23] for discussion on slack variables and examples of use).
- Penalty Terms: Constraints (e.g., flow conservation, paths disjoint-ness) become penalty terms in the objective function, i.e., these are added to the Hamiltonian with sufficiently large coefficients to ensure constraint satisfaction in low-energy solutions.
- Quadratic Terms: as QUBO allows quadratic terms as highest order polynomial, the formulation works perfectly for the L3 problem where the resiliency objective expressed in (5) is quadratic in nature.

The expression below, containing 6 terms, represents the QUBO Hamiltonian formulation for the problem:

$$\min f(x) + Bg(x) + \alpha[p(x) + q(x) + r(x) + s(x)]. \quad (7)$$

$f(x)$ and $g(x)$ are as defined previously in (2) and (5) respectively. $p(x)$ enforces flow conservation for a given path's source vertex by ensuring that it has a single active edge. $p(x)$ is defined as:

$$p(x) = \sum_{l \in T} \left[ 1 - \left(x_s^{(s,l)}\right)^2 + \left(x_s^{(s,l)} - \sum_j x_{s,j}^{(s,l)}\right)^2 \right]. \quad (8)$$

$p(x)$ is zero only when the source vertex is included in the path ($x_s^{(s,l)} = 1$) and it has exactly one active edge. Similarly, $q(x)$ enforces flow conservation for a given path's destination vertex. $q(x)$ is defined as:

$$q(x) = \sum_{l \in T} \left[ 1 - \left(x_l^{(s,l)}\right)^2 + \left(x_l^{(s,l)} - \sum_j x_{j,l}^{(s,l)}\right)^2 \right]. \quad (9)$$

$r(x)$ enforces flow conservation for a given path's intermediate vertices by ensuring that each has two active edges. $r(x)$ is defined as:

$$r(x) = \sum_{l \in T} \sum_{i \in V \setminus \{s,l\}} \left(2x_i^{(s,l)} - \sum_j x_{i,j}^{(s,l)}\right)^2. \tag{10}$$

Finally, $s(x)$ enforces the vertex disjoint paths criterion. $s(x)$ is defined as:

$$s(x) = \sum_{i \in V \setminus \{s,t_1,t_2\}} x_i^{(s,t1)} x_i^{(s,t2)}. \tag{11}$$

To conclude on the QUBO formulation, we must comment on the coefficient α in (7). The purpose of α is to assign a penalty to constraints to ensure that they are satisfied. As formulated α can be set to twice the sum of all edge weights. Because the formulation solves for two disjoint paths, such a penalty guarantees that the constraints are satisfied for the optimal (minimum energy) solution.

## 6. Quantum Approximate Optimization Algorithm (QAOA)

QAOA is a hybrid quantum-classical algorithm first introduced in [1], which presents an efficient marriage of classically computing the objective function and constraint satisfaction for a given solution while using quantum hardware to evolve the cost Hamiltonian and explore the solution space. QAOA can be viewed as a discretized version of the adiabatic evolution of a cost Hamiltonian to its lowest energy state, which is typically solved in the continuous domain on quantum annealers. The core of QAOA involves solving for the quantum state $|\gamma, \beta\rangle$ which encodes the optimal circuit parameters, defined as:

$$|\gamma, \beta\rangle = \prod_{i=1}^{p} e^{-i\beta_i H_M} e^{-i\gamma_i H_C} |s\rangle, \tag{12}$$

where $\gamma_i$ and $\beta_i$ define the strength on the $i$-th QAOA layer. Within each QAOA layer, $\gamma$ and $\beta$ define the amount of time evolution of the input state respectively under the cost Hamiltonian ($H_C$) and the mixer Hamiltonian ($H_M$). The cost Hamiltonian is the QUBO formulation of the problem expressed in (7). While the mixer Hamiltonian, which represents preference to explore more of the solution space, is usually coded as a sum of Pauli-X operators. As can be seen in (12), QAOA applies repeated unitary operations on these two Hamiltonians, up to $p$ times to an initial state, $|s\rangle$. This makes it ideal for gate-based quantum computing.

The crux of QAOA is to find the optimal value of parameters $\beta$ and $\gamma$ such that the expected value for the lowest energy state of the cost Hamiltonian is maximized, i.e. confidence in solution quality is maximized. Thus, to obtain an optimal solution to our original QUBO problem formulation, we must first solve an intermediate optimization problem to obtain $\gamma^*$ and $\beta^*$. With $\gamma^*$ and $\beta^*$, we can perform one more forward run on our quantum circuit and measure on the quantum hardware to obtain the final state $|s^*\rangle$ which will be an approximation of the true solution to the problem, where the resulting values of the qubits map to the binary decision variables $x_i^{(s,l)}$ and $x_{i,j}^{(s,l)}$. As is typically done in quantum computing, many runs are typically performed in which the output of the quantum circuit is measured and the solution with lowest energy is assumed to be the optimum.

For the QAOA implementation, we used the Classiq platform to build a high-level functional model and synthesize it to obtain the resulting gate-based quantum circuit [24]. Under the Classiq coding paradigm, objective function and constraints are all defined with traditional Python syntax. The full implementation using Classiq's Qmod language was made available on GitHub in the form of a tutorial notebook [25].

A high-level representation of the synthesized quantum circuit is shown in Fig. 2. Only one layer of the unitary cost and mixer operations are shown in Fig. 2. However, the actual implementation may use as many layers as desired—limited only by practical compute time and cost constraints. Also note that the 24 qubits required for modeling the problem for a given source vertex are well within the capabilities of quantum simulators and today's gate-based quantum computers.

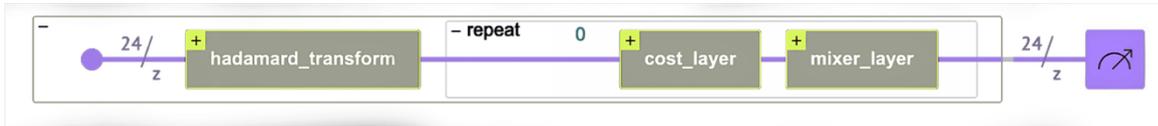

**Figure 2.** High level circuit diagram of the QAOA algorithm. Initial state of 24 qubits in $|0\rangle$ state is input to the circuit. Hadamard gate that follows prepares a superposition sate for the 24 qubits. A sequence of unitary operations that alternate between the cost Hamiltonian and the Mixer Hamiltonian is applied and repeated n times corresponding to the number of QAOA layers. The Z-gate is applied to prepare the output (classical measurement follows).

## 7. Results and Discussion

To serve as a benchmark for comparison, we first introduce the optimal solution to the problem obtained by brute force over path selection. The brute force algorithm involves the following steps:

- For each source vertex, find the set of all distinct simple paths to each of the terminal sites T1 & T2. Example for (S2,T1), the set of distinct paths is {(S2,T1), (S2,S1,T1), (S2,S3,S1,T1), (S2,S3,T2,S1,T1)}.
- A candidate solution for a source vertex combines one path from the set of paths to T1 and another path from the set of paths to T2.
- The candidate solution is checked for validity; if the two paths to T1 & T2 are not vertex-disjoint then the solution is rejected.
- Otherwise, the solution is evaluated according to cost function expressed in (1), with $B = 1000$.
- All possible combinations from the two sets are considered; the one combination with the lowest cost corresponds to the optimal solution.

By adopting brute force search over paths, the search space is significantly reduced from $2^{24}$ (searching over binary decision variables) to ~10 (to be exact, 9 for S1 and 12 for each of S2 and S3). However, such brute force approach would still suffer from combinatorial explosion as the size of the network increases since finding all distinct simple paths between given source and destination vertices in a graph is known to be NP-hard [26].

The brute force solution is shown in Fig. 3. We considered two limiting scenarios for the joint probability of link failure: independent (uncorrelated) and correlated. In both scenarios, we assume that every edge has a probability of failure of 0.1. In the first scenario, failures are assumed to be independent. Subsequently, the joint probability of failure $p(ij, mn) = p(ij) \cdot p(mn)$ is 0.01 for all combinations of edge pairs. Therefore, the solution visualized in Fig. 3a (top row) for the three secondary sites corresponds to minimum overall latency (total of 8 units). In the second scenario, it is assumed that edges (S1,S2) and (S2,T1) are 100% correlated, while all other pairs of edges remain independent with respect to link failure. The joint probability of failure becomes 0.1 for this pair. The solution visualized in Fig. 3b (bottom row) shows that the optimal path from S2 to T2 changed to avoid using the correlated edges for source S2. However, this added resiliency comes at a higher latency cost (total of 9 units).

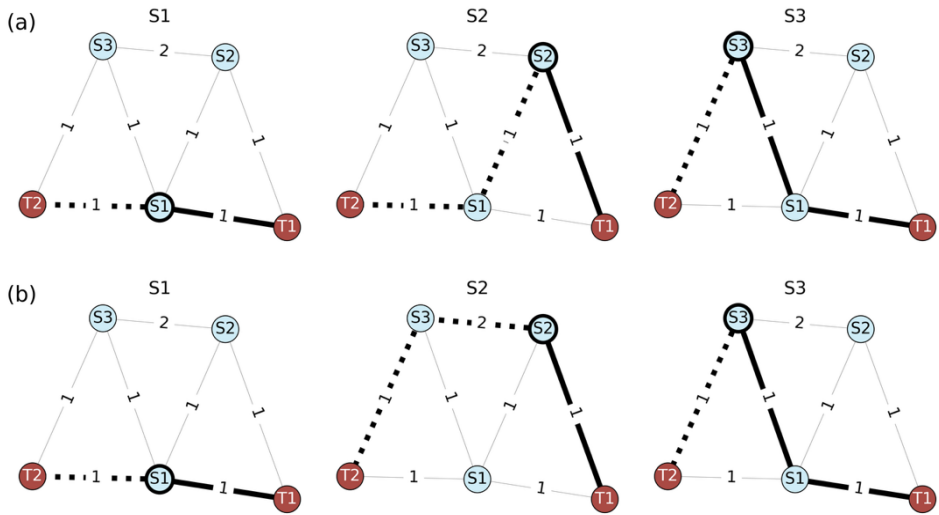

**Figure 3. Brute force solution of the problem for two scenarios. (a) Link failure is independent. In this case, the joint failure probability for any two links is 0.01. For each source (column), two disjoint paths to T1 (thick solid edges) & T2 (thick dotted edges) are highlighted. These represent the lowest latency disjoint paths. Total sum of latencies across all source-destinations is 8 units. (b) Links (S1,S2) and (S2,T1) are assumed to be 100% correlated, resulting in joint failure probability of 0.1. The solution for S2 (middle column) reconfigured one of the paths to avoid utilizing both correlated links. Total sum of latencies across all source-destinations is 9 units.**

QAOA was implemented with two variations. The first variation used 20 layers and 2480 2-qubit gates (all CX gates), executed solely on the Classiq simulator. This was done to benchmark the quantum algorithm against the brute force solution, ensuring that our modeling of the cost and constraints yields the expected results. The second variation used 8 layers and ~1000 2-qubit gates (960 and 1040 respectively for the uncorrelated and correlated link failure scenarios). This smaller circuit was optimized on the Classiq simulator for subsequent execution with the optimal parameters ($|\gamma^*, \beta^*\rangle$) on an IonQ Forte quantum computer—a system with 29 algorithmic qubits [27]. For both variations, executions on the classical simulator were done with 20000 shots, and executions on IonQ Forte were done with up to 2000 shots.

First, we introduce the results for the larger (20-layer) circuit implementation executed on the classical simulator. These are summarized in the frequency charts shown in Fig. 4, in which for each combination of joint link failure scenario and source vertex, the frequencies of occurrence for the top 10 solutions are displayed. For both uncorrelated and correlated link failures, and for all three source vertices, the solution with the highest probability matches the solution achieved by the brute force method. Furthermore, we see that the occurrence frequency drops rather quickly then levels off after the first few highest-ranking solutions. It is also observed that high ranking solutions include invalid solutions—shown as red-colored (dark) bars. This is not problematic since solutions returned by QAOA must be checked for validity against constrains (3) and (4) using classical code.

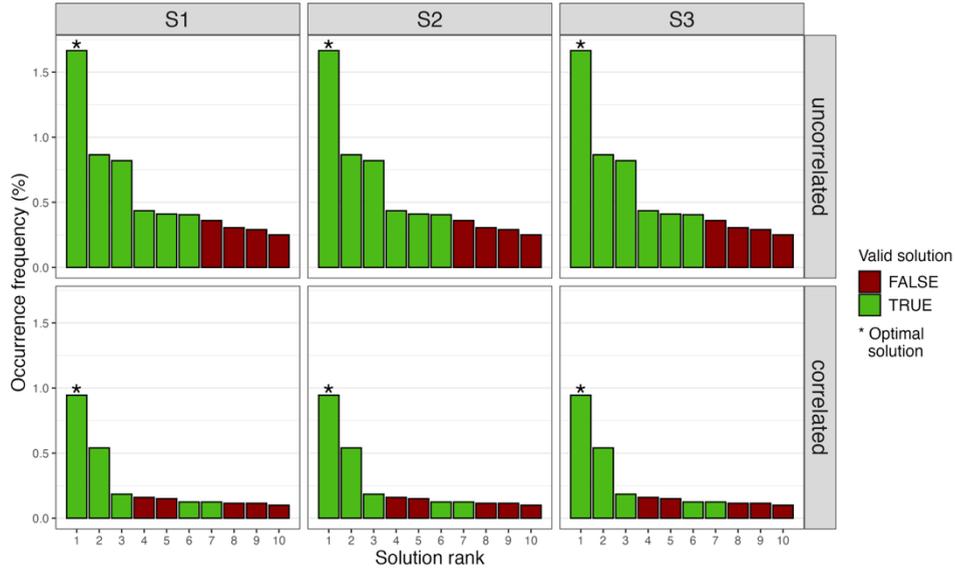

**Figure 4. Results achieved by the classical simulation of the QAOA algorithm in the form of the top 20 solutions retuned by the simulator. (Top row) Link failure is independent (uncorrelated). (Bottom row) Links (S1,S2) and (S2,T1) are assumed to be 100% correlated, resulting in joint failure probability of 0.1. In both scenarios and for all source vertices, the solution with highest frequency of occurrence corresponded to the optimal solution (asterisk on column) as determined by the brute force algorithm.**

So far, we have confirmed the feasibility of solving the L3 problem using the QAOA formulation. However, we also wanted to explore the feasibility of executing QAOA on real quantum hardware—an IonQ Forte machine. To improve the fidelity of the results when running on quantum hardware, we reduced the number of layers from 20 to 8, thus reducing the number of 2-qubit gates. Additionally, because of cost constraints, we only tested the three cases listed in Table 1—including the interesting case with S2 as source vertex, in which failures are 100% correlated for edge pair (S1,S2) and (S2,T1). In addition, test case 3 utilized IonQ's error mitigation capability based on de-biasing [28].

With regards to IonQ Forte's hardware, the dominant factor impacting results is the number of 2-qubit gates. Since the fidelity of a single 2-qubit gate is 99.6% [29], it implies that with the approximately 1000 2-qubit gates required for the problem, one may expect an overall circuit fidelity of roughly 1.8%. Combining the hardware fidelity with the probability of measuring the correct solution as achieved on the simulator, we expect a ~0.01% chance of measuring the optimal solution—equivalent to an average of 1 measurement per 10000 shots. This expected occurrence frequency is extremely low given that we ran QAOA with less than 2000 shots. However, one should view the calculation of circuit fidelity as a worst-case estimate since the true fidelity of a multi-qubit system depends in a complex way on the actual connectivity and structure of the gates within the circuit. Indeed, the IonQ Forte's results summarized in Table 1 reveal occurrence frequencies that exceed the worst-case estimate by at least an order of magnitude. The top-ranking valid solution was measured 2 times for test case 1, 5 times for test case 2, and 16 times for test case 3. Most other solutions were measured only once. Moreover, in all test cases, the top-ranking valid solution corresponded to the optimal solution. Thus, despite the somewhat low frequency of occurrence for the top solutions, this result is very encouraging as it demonstrates that the algorithm can identify optimal solutions even with current hardware limitations. The search space for the problem is immensely large ($2^{24}$) and so obtaining the optimal solution even a small fraction of the time with fewer than 2000 shots validates the QAOA algorithm for real network optimization problems. It also demonstrates the effectiveness of error mitigation using de-biasing as evident by the results of test case 3.

Table 1. Summary of test cases run on quantum hardware.

| Test Case | Source vertex | Correlated Failure Scenario | No. of shots | Error Mitigation | Optimal solution count (percentage) |
|---|---|---|---|---|---|
| 1 | S2 | Yes | 1000 | No | 2 (0.20%) |
| 2 | S1 | No | 2000 | No | 5 (0.25%) |
| 3 | S1 | No | 2000 | Yes | 16 (0.8%) |

## 8. Future Outlook

The presented work validates the formulation of the L3 latency-resilient routing problem and the feasibility of execution on quantum hardware today for a toy graph topology. Real production network topologies, however, are much larger in scale. Example, the L1 topology for a representative region in Comcast's network includes 30 vertices and 42 edges. The L3 problem formulation for such a network involves 144 binary decision variables per source site, and the corresponding QAOA 20-layer circuit has around 16k 2-qubit gates. Assuming based on the solution for the toy topology that a ~2% worst-case circuit fidelity is acceptable, implies that the 2-qubit gate fidelity must be better than 99.98% to solve the production-scale problem on quantum hardware. We should note that this is a crude estimate. The number of 2-qubit gates is also dependent on the dual link failure correlations, with more correlated pairs resulting in increased circuit complexity. The number of QAOA layers may also need to be adapted for larger circuits to achieve convergence. Adding to that, limitations on the number of qubits is another important dimension to consider as a prerequisite to scale.

We argue that the viability of quantum approaches for network optimization problems of the type presented in this paper continues to improve as quantum hardware matures. Recent advances have already demonstrated 2-qubit gate fidelities approaching 99.99% [30], a critical threshold for reducing error accumulation in multi-gate quantum circuits such as those required by QAOA. In addition, the roadmaps for several hardware vendors—including IonQ—project that 1000 algorithmic qubits are within reach by 2028 [31]. Concurrently, as quantum computing infrastructure scales and becomes more accessible, the cost per quantum circuit execution is expected to decline significantly, enabling practitioners to run algorithms with substantially higher shot counts to improve solution quality and reliability. These parallel trends in qubit count, circuit fidelity, and economic accessibility suggest that the network optimization problems formulated in this work—balancing latency and resilience in meshy topologies—will transition from theoretical demonstrations to practical deployment on quantum hardware in the near future. As this transition occurs, the QUBO/QAOA framework developed and validated in this work provides a ready pathway for telecommunications operators to leverage quantum resources for real-world path optimization at scale.

## Appendix

The appendix is dedicated to proving that NP-hardness of *min-prod* would imply NP-hardness of the L3 latency-resilient routing problem. To do this, we take an instance of *min-prod* and describe a polynomial-time reduction to the L3 problem.

Let $G$ be an undirected graph with vertices $s$ and $t$. We seek a solution to *min-prod*, i.e. a pair of internally vertex-disjoint paths $P_1$ and $P_2$ from $s$ to $t$ such that $(|P_1| + C)(|P_2| + C)$ is minimized for some given constant $C \geq 0$. Let $t_1$ and $t_2$ be arbitrary neighbors of $t$. In the terminology of Section III, we consider this an instance of the L3 latency-resilient problem with a single secondary site $s$. Taking $c_{ij}$, $d_s$ and $p(ij, mn)$ to be suitable constants, we can write $f(x) = |P^{(s,t_1)}| + |P^{(s,t_2)}|$ and $g(x) = |P^{(s,t_1)}| \cdot |P^{(s,t_2)}|$. A solution to the L3 problem consists of internally vertex-disjoint paths $P^{(s,t_1)}$ and $P^{(s,t_2)}$ such that the objective $f(x) + Bg(x)$ is minimized. Let $P_i$ denote the path $P^{(s,t_i)}$ plus the edge from $t_i$ to $t$. Scaling the objective by $B^{-1}$, adding $B^{-2}$ and setting $B = (1 + C)^{-1}$ we obtain the equivalent objective:

$$B^{-1}(f(x) + Bg(x)) + B^{-2} = (|P^{(s,t_1)}| + B^{-1})(|P^{(s,t_2)}| + B^{-1})$$
$$= (|P^{(s,t_1)}| + 1 + C)(|P^{(s,t_2)}| + 1 + C) = (|P_1| + C)(|P_2| + C)$$

Suppose that we are given a solution to the L3 problem, i.e. paths $P^{(s,t_i)}$ that minimize the objective above, for all pairs $(t_1, t_2)$ of neighbors of $t$. There are $O(n^2)$ such pairs, where $n$ is the number of vertices of $G$. Let $(t'_1, t'_2)$ be the pair for which the objective is minimized among all such pairs. This is precisely the objective for the *min-prod* problem, so the paths $P'_i = P^{(s,t'_i)}$ form a solution to *min-prod*. Hence if the L3 problem is solvable in $O(h(n))$ time, then *min-prod* is solvable in $O(n^2 h(n))$ time, thus, completing the proof.